\documentclass[review]{elsarticle}

\usepackage{lineno,hyperref}
\usepackage{graphicx}
\usepackage{color,soul}
\usepackage{natbib}
\usepackage{float}
\usepackage{rotating}
\graphicspath{{figures/}} 
\modulolinenumbers[5]
\journal{Journal of Theoretical Biology}
\DeclareMathSymbol{\mh}{\mathord}{operators}{`\-}
\usepackage{amsmath}
\usepackage[noend,linesnumbered,ruled]{algorithm2e}
\usepackage{algorithmic}
\usepackage{caption}
\usepackage{tablefootnote}

\usepackage{titlesec}
\titleformat*{\section}{\normalfont\large\bfseries}
\titleformat*{\subsection}{\normalfont\normalsize\bfseries\slshape}

\titleformat{\paragraph}[runin]{\normalfont\normalsize\bfseries}{\theparagraph}{1em}{}
\titlespacing*{\paragraph}{0pt}{0 pt}{1em}
\usepackage[section]{placeins}

\newcommand{\beginsupplement}{%
	\titleformat*{\section}{\normalfont\normalsize\bfseries}

       \setcounter{table}{0}
        \renewcommand{\thetable}{S\arabic{table}}%
        \setcounter{figure}{0}
        \renewcommand{\thefigure}{S\arabic{figure}}%
        \setcounter{section}{0}
        \renewcommand{\thesection}{S\arabic{section}}%
      }

\biboptions{authoryear}

\begin{document}
\begin{frontmatter}

\title{Modelling the effect of within-host dynamics on the diversity of a multi-strain pathogen}

\author[mainaddress]{Nefel Tellioglu}
\cortext[mycorrespondingauthor]{Corresponding author: r.chisholm@latrobe.edu.au}
\author[mainaddress,secondaryaddress]{Nicholas Geard}
\author[thirdaddress,fourthaddress]{Rebecca H. Chisholm\corref{mycorrespondingauthor}}

\address[mainaddress]{School of Computing and Information Systems, The University of Melbourne, Australia}
\address[secondaryaddress]{Department of Infectious Diseases, The University of Melbourne, Australia}
\address[thirdaddress]{Melbourne School of Population and Global Health, The University of Melbourne, Australia}
\address[fourthaddress]{Department of Mathematics and Statistics, La Trobe University, Bundoora, Australia}

\begin{abstract}
Multi-strain pathogens such as Group A \emph{Streptococcus, Streptococcus pneumoniae}, and \emph{Staphylococcus aureus} cause millions of infections each year with a substantial health burden. Control of multi-strain pathogens can be complicated by the high strain diversity often observed in endemic settings. It is not well understood how high strain diversity is maintained in populations, given that they compete with each other both directly (within an individual host) and indirectly (via host immunity). Previous modelling studies have investigated how indirect competition affects the prevalence and diversity of strains. However, these studies often make simplifying assumptions about the direct competition that occurs within hosts. Currently, little data is available to validate these assumptions, hence there is a need to clarify how sensitive model outputs are to these assumptions. In this study, we compare the dynamics of multi-strain pathogens under different assumptions about direct competition between strains using an agent-based model. We find that the assumptions made about direct competition can affect the epidemiological dynamics, particularly when there is no long-term immunity following infections and a low rate of importation of non-circulating strains. Our results suggest that while direct and indirect competition can each decrease strain diversity when they act in isolation, they may increase strain diversity when they act together. This finding highlights the importance of examining sensitivity to assumptions about strain competition. In particular, omitting consideration of direct competition can lead to inaccurate estimates of the likely effectiveness of control strategies as changes in strain diversity shift the level of direct strain competition.
\end{abstract}


\end{frontmatter}

\section{Introduction}
Multi-strain pathogens such as Group A \emph{Streptococcus, Streptococcus pneumoniae}, and \emph{Staphylococcus aureus} cause millions of infections each year \citep{obolski2018,kachroo2019}. These infections can lead to severe health conditions such as invasive infection, acute rheumatic fever and rheumatic heart disease, especially in endemic settings \citep{sanyahumbi2016}. The high strain diversity of such pathogens makes it challenging to develop a vaccine that is effective against all circulating strains. Multivalent vaccines that provide immunity against only a subset of strains may lead to reduced effectiveness via vaccine-induced serotype replacement and vaccine-induced metabolic shift \citep{weinberger2011,steer2013,  watkins2015}. 

The level of strain diversity observed in populations can vary between pathogens, populations, and over time~\citep{kucharski2016, lipsitch2009}. For example, Group A \emph{Streptococcus} maintains a higher level of strain diversity in low-income settings, such as Indigeneous communities of Australia, compared to high-income settings \citep{steer2009, mcdonald2007}. Strain diversity can be also affected by interventions \citep{obolski2018, bottomley2013}. For example, a multivalent vaccine has altered the prevalence of different metabolic genes of \emph{Streptococcus pneumoniae} in the United States \citep{watkins2015}. The ecological mechanisms underlying the distribution and maintenance of these varying levels of strain diversity are not well understood \citep{smeesters2010,pinotti2019}. Strain diversity is likely to be influenced by the competitive interactions between strains that occur both directly (within-hosts) and indirectly (between-hosts) \citep{mideo2008, alizon2011, bashey2015, watkins2015, georgieva2019, smeesters2009}. For example, within an individual host, strains with the same metabolic profile may compete with each other for the same substrates \citep{mideo2009}; or strains may produce toxins, such as bacteriocins, which inhibits the growth of other strains \citep{bashey2015}. At the between-host level, a strain which infected a host previously may prevent other strains with the same antigenic type to infect the same host whose immune system has memory after infection  \citep{gog2002}.

Previous modelling studies have investigated how indirect competition affects strain diversity \citep[\emph{e.g.}][]{alizon2013}. \cite{gog2002} showed that when the immune response to an infection by one strain provides some level of immunity against other strains (\emph{cross immunity}), multiple strains with varying serotypes can only co-exist in the population when the duration of immunity is sufficiently long, otherwise at most only one group of strains with similar serotypes can co-exist. \cite{chisholm2020} showed that diversity may also depend non-monotonically on transmissibility and can be maximized by an intermediate level of transmissibility. These models tend to make varied assumptions about the direct competition that occurs within hosts. For example, some models assume that there is no limit on how many antigenically different strains can co-infect a fully susceptible individual host at the same time \citep{dawes2002, flasche2013}. While other models assume that co-infection is not possible and that one strain will always outcompete other strains \citep{abu2004,aquino2015, alizon2013}. Some models incorporate the effects of competition within hosts in a phenomenological way by, for example, assuming there is increasing resistance to co-infection as additional strains co-infect the host \citep{buckee2004,chisholm2020, bianco2009}. Other models incorporate more biologically realistic representation of within-host competition. For example, in the model proposed by \cite{watkins2015} strains were classified according to their antigenic and metabolic type, and it was assumed that infection could not occur if a host was currently infected by a strain with the same metabolic type. There is often little immunological (host level) and strain-specific metabolic (strain level) data to support these assumptions about direct competition \citep{watkins2015, chisholm2020}. Given this uncertainty, multi-strain models with different assumptions of direct competition may produce different epidemiological dynamics. Therefore, there is a need to clarify how sensitive the model outputs are to these assumptions.

In this study, we develop a flexible agent-based model to compare the dynamics of multi-strain pathogens under different assumptions about direct competition of strains. We characterise the sensitivity of model outputs to the assumptions made about direct competition, and describe how this sensitivity varies with the duration of immunity, pathogen transmissibility, and host demographic characteristics. We find that assumptions about direct competition have the greatest impact on epidemiological dynamics when the duration of immunity is short and disease transmissibility is low. As studies of Group A \emph{Streptococcus} and \emph{Streptococcus pneumoniae} consider antigen specific immune response \citep{denny1957, lancefield1959, weinberger2008, chisholm2021} with unknown immunity durations \citep{hysmith2017, weiser2018},  these pathogens may have such characteristics, and therefore, direct competition can affect their epidemiological dynamics. Therefore, understanding model sensitivity to these characteristics is important, particularly when models are being used to support public health decisions.

\section{Methods}
Our stochastic agent-based model~\footnote{Code is available online at \url{https://github.com/nefeltellioglu/multistrain}} describes the transmission of a hypothetical multi-strain pathogen in a host population where agents represent hosts. In this section, we describe the host population, how we represent strains and their interactions, transmission events and  the design of our experimental studies. 
\subsection{Host population}

The host population has a constant size $N$ and an underlying fixed social contact network which we construct using an Erdős-Réyni model with mean degree $c$ (where $c$ is proportional to the mean daily number of social contacts between agents). The state of agents is updated at discrete time steps. Agents are characterised by their age, the set of strains they are currently infected with, and the list of strains they are immune to. 

New individuals are introduced into the population via birth-death and migration processes. When an agent dies, this agent is replaced by a fully susceptible agent with age 0 is introduced to the population so that the population size $N$ remains constant. At each time point, individuals emigrate from the population with a probability that depends on the per capita migration rate,  $\alpha$, which corresponds to the emigration of, on average, $\alpha N$ randomly selected agents at each time point. When an agent emigrates, it is replaced by a newly generated (migrant) agent. We assume that the prevalence of infection in migrants is $s$ and that infected migrants are infected by 1 strain chosen uniformly at random.

We also assume that external populations have similar epidemiological characteristics to the modelled population. Therefore, migrants are assumed to have a similar immune history to the population which we implement by setting the immune history of new migrants to that of a randomly selected individual from the population.  

\subsection{Strain representation and interaction}
We characterise a strain $(i,j)$ by its antigenic type (AT) $i$, and metabolic type (MT) $j$, where $i\in\{1,2, \cdots,A\}$ and $j\in\{1,2, \cdots,M\}$, where $A$ is the number of antigenic types and $M$ is the number of metabolic types. Following \cite{watkins2015}, we assume that a strain's AT controls how they compete indirectly via host immunity, while their MT determines how they compete directly for resources within the host (described in further detail below). We characterise the transmissibility of the pathogen by the \emph{basic reproduction number}, $R_0$, which is the expected number of secondary infections from a single infected host in a fully susceptible host population. We assume that all strains have equal transmissibility or duration of infection. Therefore, each strain has an identical strain-specific basic reproduction number $R_0$ which is equal to the overall basic reproduction number for the pathogen:

\[R_0=\frac{c * q}{(1/ \gamma) + d + \alpha}\]where $c$ is the average number of contacts, $q$ is the probability of transmission in a contact, $\gamma$ is the mean duration of infection, d is the average death rate, $\alpha$ is the per capita migration rate. 

\begin{figure}[!htb]
\centering
\includegraphics[width=1\textwidth]{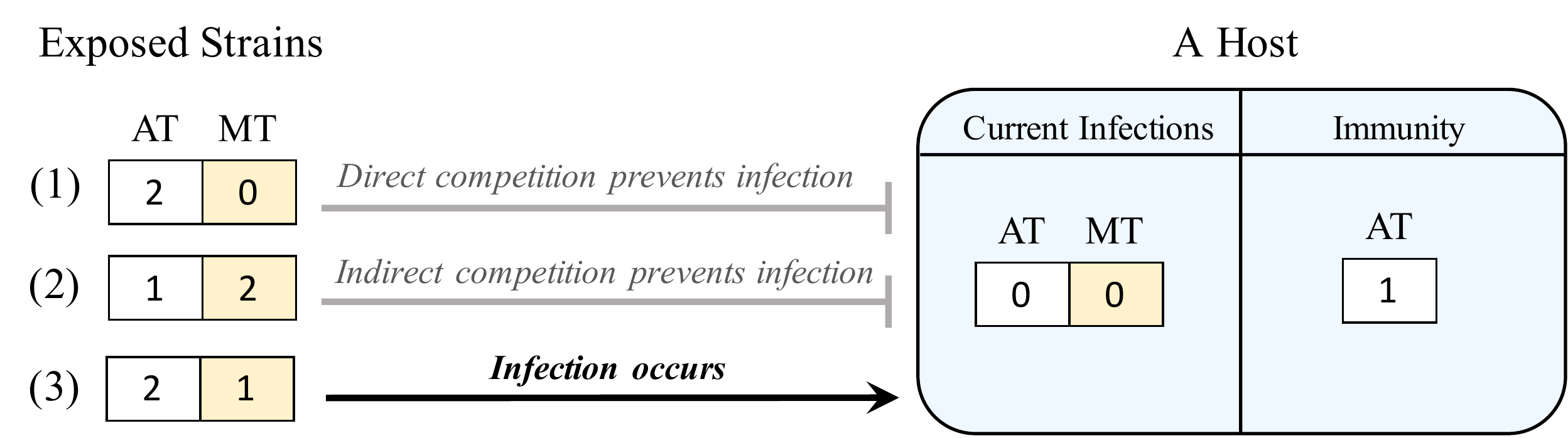}
\caption{Exposed strains can only infect the host if there are no direct or indirect competitive interactions. Exemplar exposure/transmission scenarios are shown where a host who is currently infected by strain (0,0) and immune to AT 1 is exposed to three different strains. While the first two strains cannot infect the host (due to direct and indirect competition, respectively), the third strain can co-infect the host.}
\label{fig:at_mt_example}\end{figure}

\paragraph*{Indirect Competition}
After clearing an infection with a strain with AT $i$, a host develops immunity to all strains with the same AT $i$ (antigen-specific immunity). We assume that the antigen-specific immunity fully protects the host from further infection by strains with the same AT $i$, so that a strain with AT $i$ can neither infect a host with immunity to AT $i$ nor infect a host who is currently infected by a strain with AT $i$ (Figure \ref{fig:at_mt_example}). The model does not allow for cross-strain protective immunity.

\paragraph*{Direct Competition}
Direct competition among strains that co-infect a host can occur due to exploitation of the same resources, known as \emph{metabolic-type competition}; or due to production of toxic agents, known as \emph{allelopathic interference}  within a host \citep{bashey2015}. Our model represents metabolic-type competition. We assign a metabolic type to each strain, which indicates which host resources are necessary for that strain's survival \citep{watkins2015, watkins2016}. We assume that strains sharing the same MT cannot co-infect a host at the same time, as they would then be in competition for the same resources. We also assume that the strain that infects a host first cannot be displaced by another strain due to having a competitive advantage of occupying a niche first, known as \emph{priority interference} \citep{bashey2015}. Therefore, if an infected host is exposed to a new strain, transmission is only possible if the new strain has an MT that is different from any of the strains currently infecting the host (Figure \ref{fig:at_mt_example}). Our model does not explicitly present the competitive dynamics that occur within the host but just captures the outcome of the competition.

By changing the number of MTs, $M$, in the model, we can vary the level of direct competition between strains. When $M = 1$, only one strain can infect a host at a given time (high direct competition).  As $M\to\infty$, co-infection depends only on host immunity and there is no direct competition within hosts.

\subsection{Multi-strain transmission model}
Transmission can occur between agents that share an edge of the contact network (\emph{neighbours}). At each time step, each infected host has infectious contacts with their neighbours with a probability of $q$. Then, every agent checks whether or not it (i) is exposed to new strains, (ii) will recover from its infections, (iii) will lose some of its antigen-specific immunity, (iv) will migrate or die. Based on this information, agents may become infected, recover, lose their immunity, migrate, and/or die. At the end of each time step $t$, a set of emigrated agents are removed and replaced by a set of newly generated migrant agents. The pseudo code of the model is presented in Figure \ref{fig:pseudocode}. 

We assume that a co-infected host can transmit at most one of the strains it is carrying at the time of contact with a susceptible host, and a susceptible host can be infected by at most one of the strains that they are exposed to during a single time step. We assume that the duration of infection and immunity follow Gamma distributions \citep{lloyd2001} with mean $\gamma$ and with mean $\mu$, respectively. For the duration of infection, we use shape parameter 3 to generate a wide range of values around $\gamma$. For the duration of immunity, we use shape parameter 20 to generate a narrow range of values around $\mu$ so that every recovered person will be immune for at least period of time when there is immunity following infection. For example, a Gamma distribution for immunity duration with an average of 15 weeks results in a range of 6 weeks - 27 weeks. Infection and immunity distributions are presented in Figure \ref{fig:inf_imm_dist}.


\renewcommand{\thealgocf}{}

\noindent\makebox[\textwidth][c]{
\begin{minipage}{1.3\textwidth}

\SetAlFnt{\footnotesize}
\SetArgSty{textnormal}

\begin{algorithm}[H]
	\SetAlgorithmName{Algorithm}{Algorithm}{}

    Generate Erdős-Réyni network with $N$ nodes and mean degree $c$\;
     \lFor{each node $k\in\{1,2, \cdots,N\}$}{create $a_k$, a fully susceptible agent (host)}
     \lFor{each strain (i,j) $i\in\{1,2, \cdots,A\}$ and $j\in\{1,2, \cdots,M\}$}{select $I$ number of agents susceptible to strain (i,j) and infect them with strain (i,j)}
    
    \While{model tick $ t <= T$ }
      {
        Generate a shuffled list, $L(t)$, of $N$ agents\;
        \For{each infected agent $a_k\in L(t)$}
        {
        	   Randomly select one of the strains (i,j), that $a_k$ is infected by\;
	    \For{each neighbours of $a_k$, $a_n$, who are metabolically and antigenically susceptible to the strain (i,j)}{
	    Add strain (i,j) to the $a_n$.ExposedList with a probability of $q$\;
	    }
         }
	\For{each agent $a_k\in L(t)$}
        {   
        	   \If{ $a_k$.ExposedList $ \neq  \emptyset $}{
		Randomly select one of the strains, (i,j), from $a_k$.ExposedList and remove the others from $a_k.$ExposedList\;
		}  	
           \For{each strain (i,j) in $a_k$.InfectionList}{
	   	\leIf{$a_k$.InfectionTimer.(i,j)   $>$  0}{
		$a_k$.InfectionTimer.(i,j)  -= 1\;}{
		set $a_k$.WillRecoveredList.(i,j) as True}
	     } 
	     \For{each strain (i,j) in $a_k$.ImmuneList}{
		\leIf{$a_k$.ImmuneTimer.(i,j)   $>$  0}{
			$a_k$.ImmuneTimer.(i,j)  -= 1\;}{
			set $a_k$.WillSusceptibleList.(i,j) as True}}       
        	   \If{rand(0,1) $<=$ MigrationRate, $\alpha$,}{
	     set $a_k$.WillMigrate as True and \emph{NumberofEmigrated} += 1}
	    \ElseIf{$a_k$.birthday $==$ ($t \mod$\emph{t\_per\_year}) }{ $a_k$.Age  += 1\;
	     set $a_k$.WillDie as True with a probability of DeathRates.($a_k$.Age)\;
	    }
	   
	  }
	 \For{each agent $a_k\in L(t)$}
           { 
             \lIf{$a_k$.ExposedList $ \neq  \emptyset$}{add (i,j) to $a_k$.InfectionList, generate a value for $a_k$.InfectionTimer.(i,j), and set $a_k$.ExposedList $\emptyset$}
             \lFor{each strain (i,j) in $a_k$.WillRecoveredList \textbf{if}  $a_k$.WillRecoveredList.(i,j)} {add (i,j) to $a_k$.ImmuneList, generate a value for $a_k$.ImmuneTimer.(i,j) and set $a_k$.WillRecoveredList.(i,j) False}
             \lFor{each strain (i,j) in $a_k$.WillSusceptibleList \textbf{if}  $a_k$.WillSusceptibleList.(i,j)} {remove (i,j) from $a_k$.ImmuneList and set $a_k$.WillSusceptibleList.(i,j) False}
            \lIf{$a_k$.WillMigrate}{remove $a_k$ from agents} 
            \lIf{$a_k$.WillDie }{remove $a_k$ from population and generate a newborn fully susceptible agent}
	    }
	    Generate \emph{NumberofEmigrated} agents and set \emph{NumberofEmigrated} as zero\; 
    
    }
\end{algorithm}
\captionof{figure}{Pseudo code of our stochastic agent-based model}
\label{fig:pseudocode}

\end{minipage}
}

\subsection{Study design}

Our goal is to understand how sensitive model outputs are to assumptions about direct competition, and whether this sensitivity is affected by the assumed level of indirect competition in the model. We explore model sensitivity to direct competition assumptions by varying the number of MTs, $M$. The level of indirect competition in the model is altered by either considering different durations of immunity (ranging from 0 to lifelong), or different values of basic reproduction number  $R_0$  (ranging from 1 to 4).

The values we consider for $M$ are $M = 1,2,3,$ and $\infty$ which represent, respectively,
\begin{itemize}
  \item \textbf{High Direct Competition:} There is only one MT in the model ($M = 1$) which causes all strains to directly compete for resources. In this scenario, only one strain can infect a host at a given time. 
  \item  \textbf{Medium Direct Competition:} There are two MTs in the model ($M = 2$) where strains sharing the same MT compete directly for resources. In this scenario, up to two strains with different MTs can infect a host at a given time. 
   \item  \textbf{Low Direct Competition:} There are three MTs in the model ($M = 3$) where strains sharing the same MT directly compete for resources. In this scenario,up to three strains with different MTs can infect a host at a given time. 
   \item  \textbf{No Direct Competition:} There are no MTs in the model ($M\to\infty$), meaning that strains do not compete directly for resources. In this scenario, co-infection only depends on host immunity. A fully susceptible host can be infected by up to $A$ strains at a given time.
\end{itemize}

We repeat this analysis (1) in a population without migration to determine how strain re-introduction through migration affects the sensitivity of the model to the level of direct competition, and (2) in a population without migration and birth-death processes to determine how introduction of fully susceptible newborn individuals affects the sensitivity of the model to the level of direct competition. 

We parameterise our model population using the Australian mortality rate and population by age data \citep{absage,abs2020}. The values of other model parameters are shown in Table \ref{table:parameters}. We initialise our simulation with all possible combinations of AT-MT strains present in the host population.  Each simulation is run for 150 years so that the infection dynamics reach a dynamic equilibrium.  We run 10 realisations of each indirect and direct competition scenario. For each scenario, we report the following model outputs which represent strain diversity, transmissibility, and reliance on the re-introduction of strains to maintain diversity: 

\begin{itemize}
  \item The period diversity of ATs: It is a measure of the number and evenness of ATs in circulation in the population over a period of time. It demonstrates how many antigenic types circulating in the population as well as the evenness among the number of infections by each AT in a given time interval $u$ \citep{smeesters2009, chisholm2020}. We use Simpson’s reciprocal index to calculate period diversity, $D(t)$:

\[D(t)=\frac{M(t;u)(M(t;u)-1)}{\sum_{j}m_{j}(t;u)(m_{j}(t;u)-1))}\]

where $m_{j}(t;u)$ is the number of infections of AT in the population in the time interval of $[t-u, t]$, and $M(t;u)$ is the total number of infections in the population in the time interval of $[t-u, t]$. 

  \item The distribution of the effective reproduction number of ATs, , R\textsubscript{eff}(t): it is a measure of the transmissibility of ATs. For each AT \textit{j}, we calculate the mean number of secondary infections generated by infectious agents who recover from an infection with AT \textit{j} at time step \textit{t}, R\textsubscript{eff,j}(t). The distribution R\textsubscript{eff}(t) is then defined to be the set of all R\textsubscript{eff,j}(t) for $j\in\{1,2, \cdots,M\}$.

  \item The final turnover proportion of ATs: It quantifies the fraction of the observed AT diversity that derives from the importation of ATs which do not exist in the population at the time of importation. It represents the degree of AT diversity that is reliant on the re-introduction of strains via host migration. We calculate the turnover proportion by dividing the accumulative number of importations of non-circulating ATs by the accumulative number of AT importations at the end of simulation. The turnover proportion of ATs is only calculated in scenarios with migration. 
\end{itemize}

\begin{table}[!htb]

\caption[Parameter values of the model]{Parameter values of the model.}
\centering
\resizebox{1\columnwidth}{!}{%
\begin{tabular}{|p{3em}| p{20em} |p{5em}|p{7em}|}
\hline
Symbol & Description &Base Value &Other Values \\ \hline
$N$ & Population size & 2500 & -\\
$d$ & Average death rate (per year) & 0.0033 & 0 \\
$\alpha$ & Migration rate (per week) & 0.0001 & 0 \\
$s$ & Probability of a migrant being infected & 0.1 & - \\
$q$ & Probability of transmission in a contact & 0.05 & - \\
$c$ & Mean number of edges in the network (Mean number of weekly contacts per host) & 10&5.5, 6.5, 7.5, 12.5, 15, 20 \\
$\gamma$ & Mean duration of infection (weeks) & 4 & -\\
$\mu$ & Mean duration of immunity (weeks) & 0, 15, 40, Lifelong & 30 \textsuperscript{1}\\
$R_0$& Basic Reproduction Number  \textsuperscript{2}& 1.97 & 1.08, 1.28, 1.48, 2.46, 2.96, 3.94\\
$A$ & Number of antigenic types & 9 & - \\
$M$ & Number of metabolic types & 0, 1, 2, 3& - \\
$I$ & Initial number of hosts infected with strains with each antigenic type & 15 & -\\
- & Initial number of hosts immune to each antigenic type & 0 & - \\
$u$ & Time interval used in period diversity (years) & 10 & - \\
\hline
\end{tabular}
}
\begin{flushleft} 
\footnotesize
\textsuperscript{1} In demographic settings where there is (1) no migration and (2) no birth-death and migration processes, we consider 30 weeks as medium immunity duration because the effect of direct competition with an immunity duration of 30 weeks is more clearly observed than the one with immunity duration of 40 weeks when there is no migration.

\textsuperscript{2} In the results section, we use the rounded $R_0$ values when we present simulation results.
\end{flushleft}
\label{table:parameters}
\end{table}

\section{Results}

\subsection{The impact of direct competition on strain diversity varies with the duration of immunity}
We begin by exploring the sensitivity of model outputs to the number of MTs, \textit{M}, under different levels of indirect competition, which we control by varying the duration of immunity while keeping $R_0$ fixed at $2$. Here, longer immunity duration scenarios equate to higher levels of indirect competition, and higher levels of \textit{M} equate to scenarios with lower levels of direct competition.

\begin{figure}[!htb]
\centering
\includegraphics[width=0.9\textwidth]{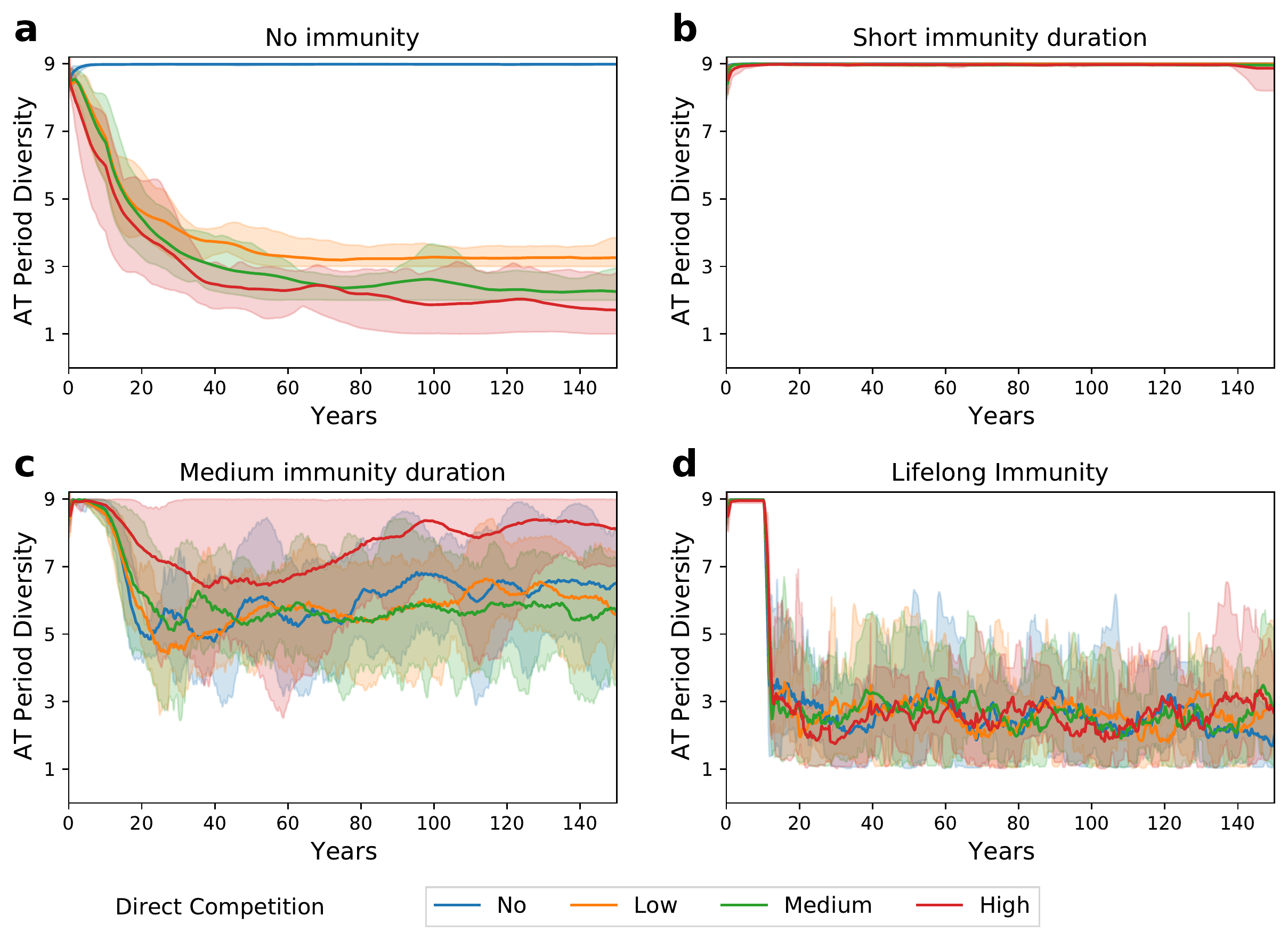}
\caption{Direct competition can increase or decrease strain diversity depending on the duration of immunity. Mean values and $2.5\% - 97.5\%$ quantiles of period diversity of circulating antigenic types under different immunity durations are shown: (a) no immunity duration,  (b) short immunity duration (15 weeks), (c) medium immunity duration (40 weeks), (d) lifelong immunity. Levels of direct competition represent scenarios where there are no MTs (no), three MTs (low),  two MTs (medium), and one MT (high). Here, $R_0 \approx 2$ ($c = 10$), $\gamma = 4$ weeks, $\alpha = 0.0001$.}
\label{fig:atdiversity}\end{figure}

We observe that changing the level of indirect competition via the duration of immunity can alter the impact of direct competition on strain diversity significantly. For example, when there is no immunity following infections, there is a clear distinction between scenarios with no direct competition and scenarios with some competition ($M = 1, 2, 3$) (Figure \ref{fig:atdiversity}.a). Specifically, we see all ATs in circulation when there is no direct competition. As the level of direct competition increases from zero to a low level ($M = 3$), the endemic diversity decreases to just over $3$, and decreases further as the level of direct competition further increased ($M = 2, 1$). This distinction is not apparent in the short immunity scenarios (Figure \ref{fig:atdiversity}.b), where there appears to be no sensitivity of endemic diversity to the number of MTs, \textit{M}, for this particular value of $R_0$. When there is a medium immunity duration (medium indirect competition, in Figure \ref{fig:atdiversity}.c),increasing the level of direct competition has the opposite effect as that observed with no immunity scenarios. Specifically, higher endemic strain diversity is observed in scenarios with higher levels of direct competition. Finally, in scenarios with lifelong immunity (high indirect competition, in Figure \ref{fig:atdiversity}.d), there is no observed effect of direct competition on endemic diversity, suggesting that indirect competition is dominating the transmission dynamics. 

\begin{figure}[!htb]
\centering
\includegraphics[width=0.65\textwidth]{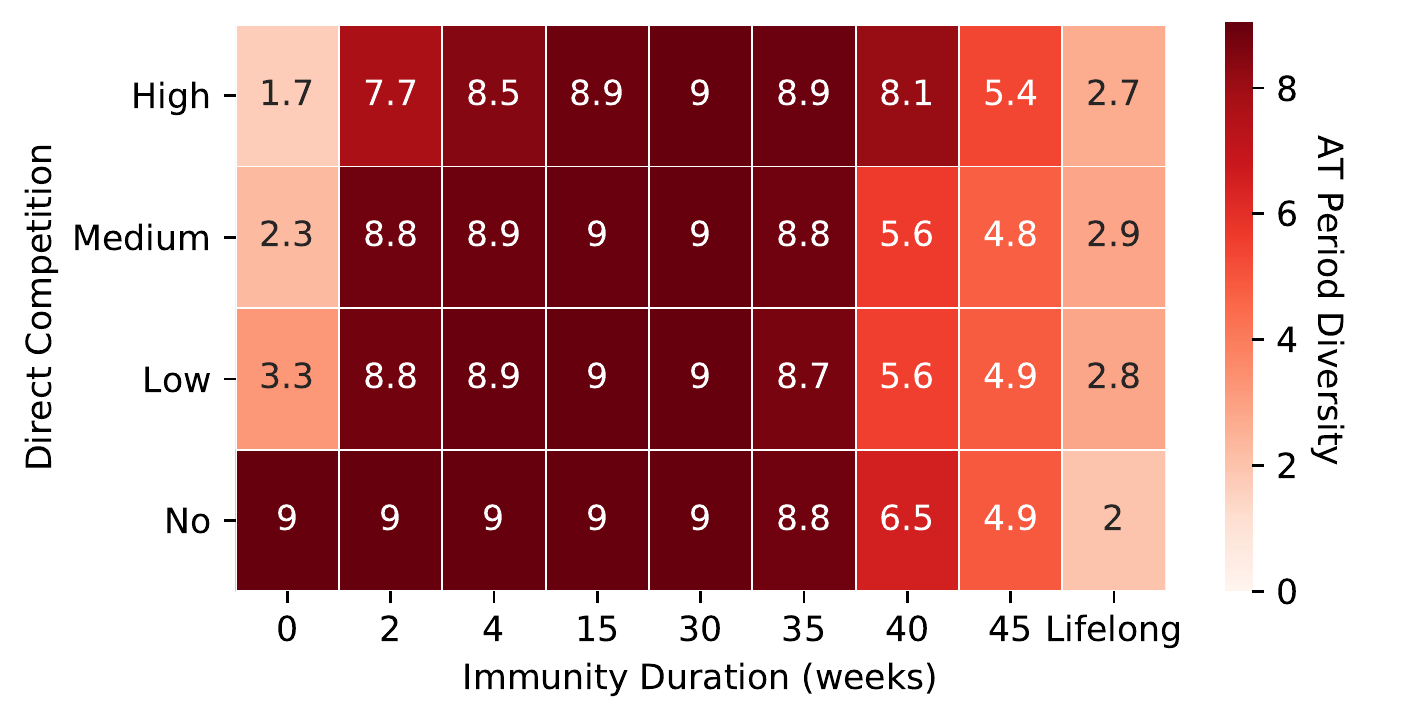}
\caption{ The effect of direct competition on strain diversity depends on the duration of immunity. Mean period diversity of antigenic types at the end of the simulations under different immunity durations are shown. Levels of direct competition represent scenarios where there are no MTs (no), three MTs (low),  two MTs (medium), and one MT (high). Here, $R_0 \approx 2$ ($c = 10$), $\gamma = 4$ weeks, $\alpha = 0.0001$.}
\label{fig:atsummary}\end{figure}

In Figure \ref{fig:atsummary}, we show results for the endemic strain diversity values (calculated at the end of the simulations) under these and additional durations of immunity. We observe that higher levels of direct competition results in lower levels of endemic strain diversity when the duration of immunity is less than 4 weeks. For immunity durations $> 4$ and $< 40$ weeks, endemic duration is insensitive to the level of direct competition. In scenarios where the duration of immunity is 40 weeks, higher levels of direct competition lead to higher endemic diversity. We analyse this changing effect of direct competition on strain diversity in Section \ref{DirectcompetitionReff}.

\subsection{The impact of direct competition on the transmissibility of antigenic types (R\textsubscript{eff}) varies with the duration of immunity} \label{DirectcompetitionReff} 

\begin{figure}[!htb]
\centering
\includegraphics[width=0.9\textwidth]{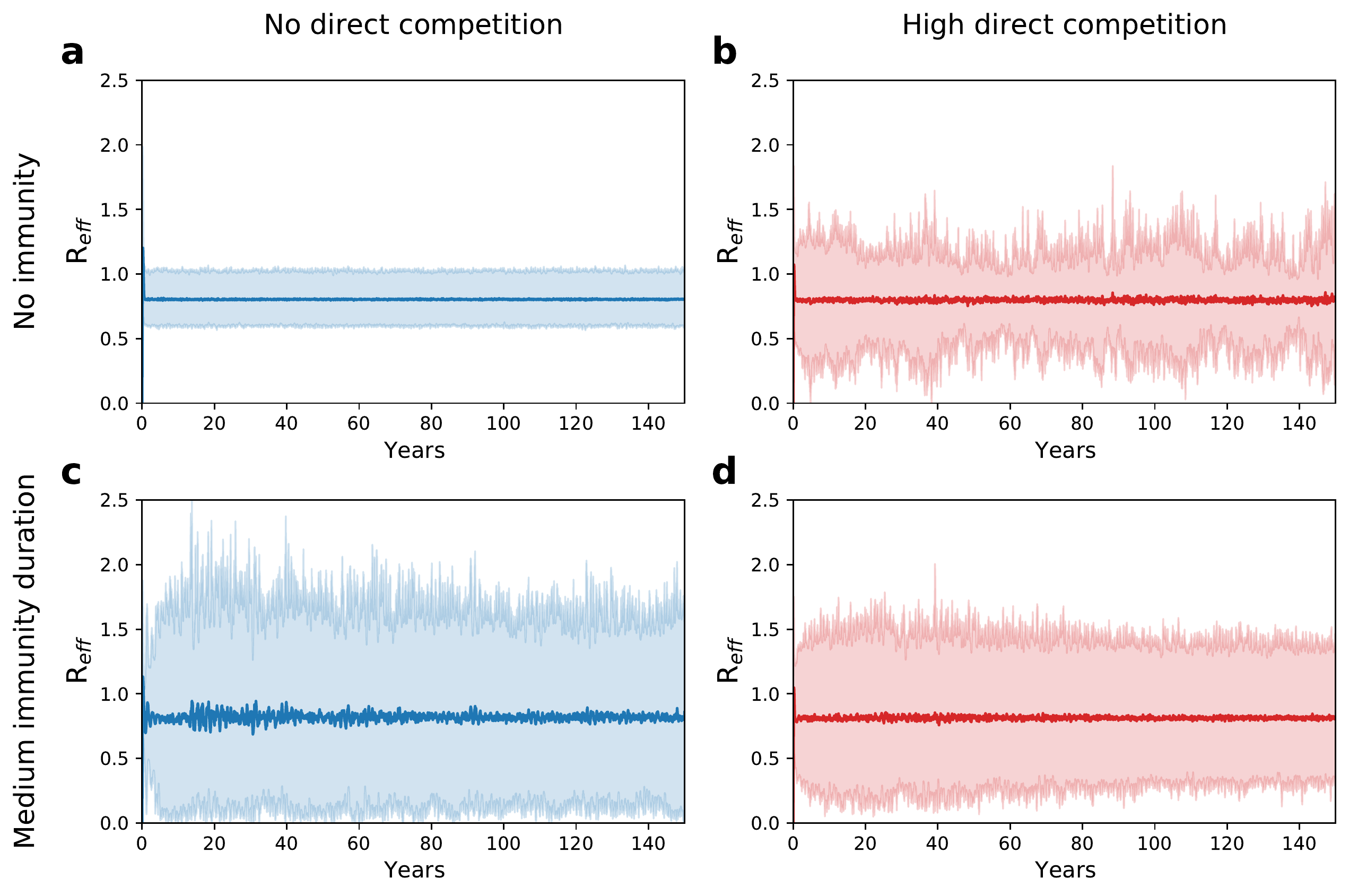}

\caption{The amplitude of oscillations of R\textsubscript{eff} of circulating antigenic types depends on both the level of direct competition and duration of immunity. Mean values and $2.5\% - 97.5\%$ quantiles of R\textsubscript{eff} of circulating antigenic types are shown. First row (a and b) is the results of models with no immunity, second row (c and d) is with medium immunity duration (40 weeks); first column (a and c) is the results of models with no direct competition (blue) and second column (b and d) is the results of models with high direct competition (red). High direct competition represent one MT and no direct competition represents no MTs. Here, $R_0 \approx 2$ ($c = 10$), $\gamma = 4$ weeks, $\alpha = 0.0001$.}
\label{fig:atreff}\end{figure}

In this section, we aim to understand why higher levels of direct competition lead to higher levels of endemic diversity when immunity is long-lasting, but lower levels of endemic diversity when immunity is short-lived. We compare R\textsubscript{eff} distributions (Figure \ref{fig:atreff}), and the final turnover proportion of ATs (Table \ref{table:oscillations2}) between scenarios with “No Direct Competition” and “High Direct Competition”, and when there is no immunity and when the immunity duration is set to 40 weeks (medium duration immunity).

We observe that the assumed level of direct competition affects both R\textsubscript{eff} distribution and the final turnover proportion of ATs. In scenarios with no immunity following infection (Figure \ref{fig:atreff}.a and \ref{fig:atreff}.b), the introduction of direct competition leads to a significant increase in the variability of R\textsubscript{eff}(t), around the mean (which is just below 1), indicating that a larger proportion of ATs are vulnerable to stochastic extinction since their individual R\textsubscript{eff} will be less than 1. In case of the medium-term immunity scenarios, the introduction of direct competition has the opposite effect - the variability of R\textsubscript{eff} around the mean is reduced, allowing more ATs to co-circulate (Figure \ref{fig:atreff}.c). This behaviour is also reflected in the final turnover proportion (Table \ref{table:oscillations2}). When there is no immunity following infection, the introduction of direct competition increases the final turnover proportion of ATs, indicating that the maintenance of strain diversity is more reliant on the re-introduction of strains via host migration compared to the scenarios without direct competition. When there is medium immunity after infection, the introduction of direct competition decreases the final turnover proportion of ATs indicating the maintenance of strain diversity is less reliant on the re-introduction of strains via host migration, compared to scenarios with no direct competition.

\begin{table}[!htb]
\caption{Mean and $[2.5\% - 97.5\%]$ quantiles of \emph{turnover proportion}, the fraction of the observed AT diversity derives from the importation of ATs. 
}
\centering
\resizebox{0.8\columnwidth}{!}{%
\begin{tabular}{c|c|c|}
\cline{2-3}
\multicolumn{1}{l|}{} & \textbf{No Direct Competition} & \textbf{High Direct Competition} \\ \hline
\multicolumn{1}{|c|}{\textbf{No Immunity}} & 0 {[}0 - 0{]} & 0.60 {[}0.53 - 0.68{]} \\ \hline
\multicolumn{1}{|c|}{\textbf{Medium Immunity}} & 0.46 {[}0.38 - 0.56{]} & 0.19 {[}0.03 - 0.40{]} \\ \hline
\end{tabular}
}
\label{table:oscillations2}
\end{table}

Now, we explore the sensitivity of model outputs to the number of MTs, \textit{M}, under different levels of indirect competition which we control by varying the basic reproduction number $R_0$ only, and keeping the duration of immunity fixed at either 0 weeks, 15 weeks (short immunity duration), 40 weeks (medium immunity duration), or lifelong. Here, higher values of $R_0$ equate to higher levels of indirect competition (when the duration of immunity $> 0$).

\subsection{The effect of direct competition on strain diversity varies with the disease transmissibility ($R_0$)}
\begin{figure}[!htb]
\centering
\includegraphics[width=1\textwidth]{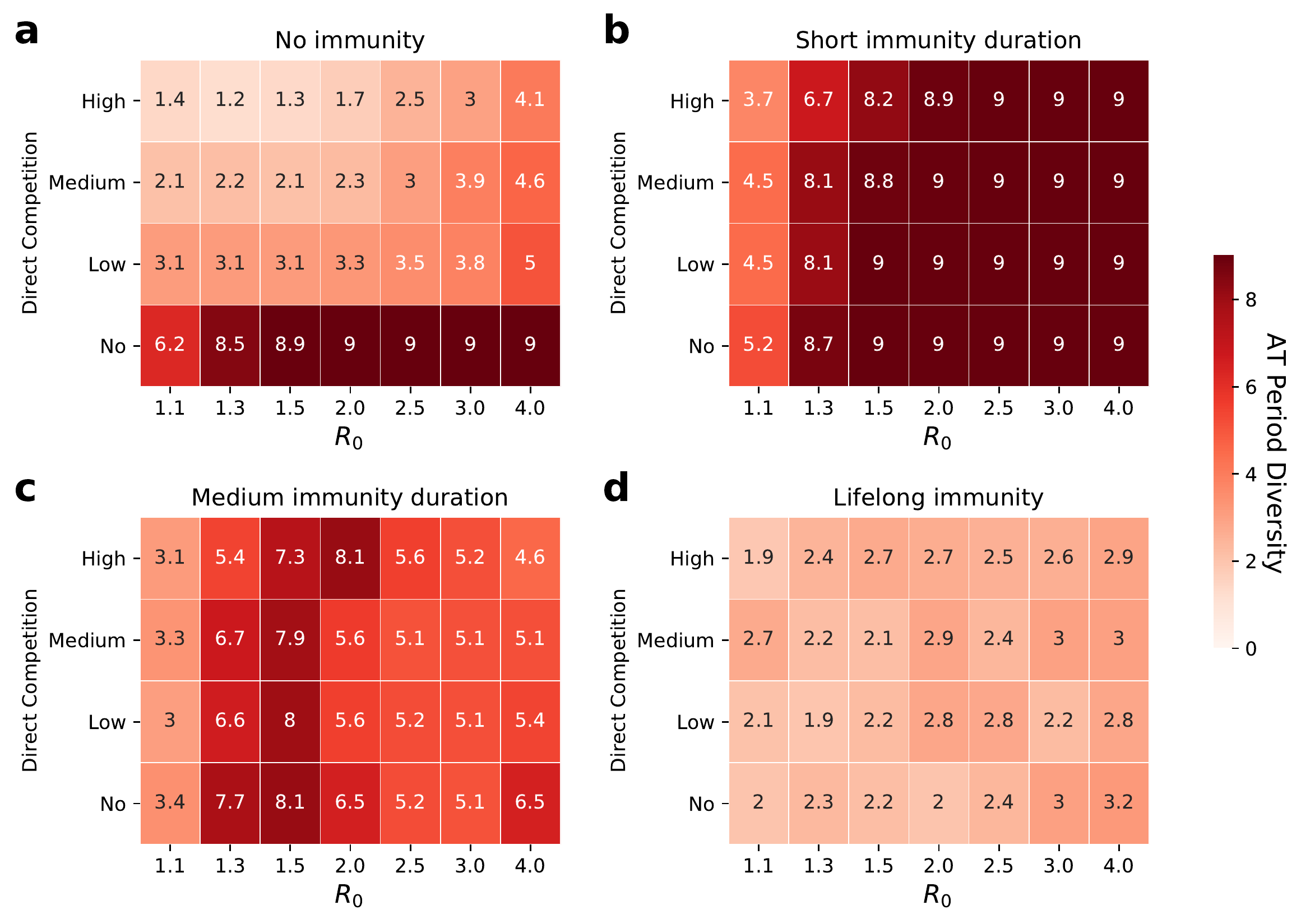}

\caption{The effect of direct competition on strain diversity depends on disease transmissibility, $R_0$. Mean period diversity of antigenic types at the end of the simulations under different $R_0$ are shown. The immunity levels are (a) no immunity, (b) 15 weeks, (c) 40 weeks, (d) lifelong immunity. Levels of direct competition represent scenarios where there are no MTs (no), three MTs (low),  two MTs (medium), and one MT (high). Here, $\gamma = 4$ weeks, $\alpha = 0.0001$.}
\label{fig:atsummaryr0}\end{figure}

Generally, we find that when the duration of immunity is sufficiently short, the endemic diversity of strains in scenarios with higher $R_0$  values is less sensitive to assumptions about direct competition compared to lower $R_0$ scenarios (Figure \ref{fig:atsummaryr0}.a-b). In these scenarios, higher levels of direct competition generally correspond to lower levels of endemic diversity. 
In medium duration of immunity scenarios, $R_0$ determines whether higher levels of direct competition correspond to higher of lower levels of endemic diversity. When $R_0$ is sufficiently small ($R_0 < 2$) or sufficiently large ($R_0 > 4$), lower levels of endemic diversity correspond to scenarios with higher levels of direct competition. Otherwise, increasing levels of direct competition lead to lower levels of endemic diversity. In scenarios with lifelong immunity, there is no sensitivities of endemic diversity to either $R_0$ or the level of direct competition. Therefore, we conclude that the effect of direct competition on strain diversity depends not only on the duration of immunity acquired following infections but also on the value of $R_0$.

\subsection{Host migration reduces the effect of direct competition on strain diversity}
\begin{figure}[!htb]
\centering
\makebox[\textwidth][c]{\includegraphics[width=1.3\textwidth]{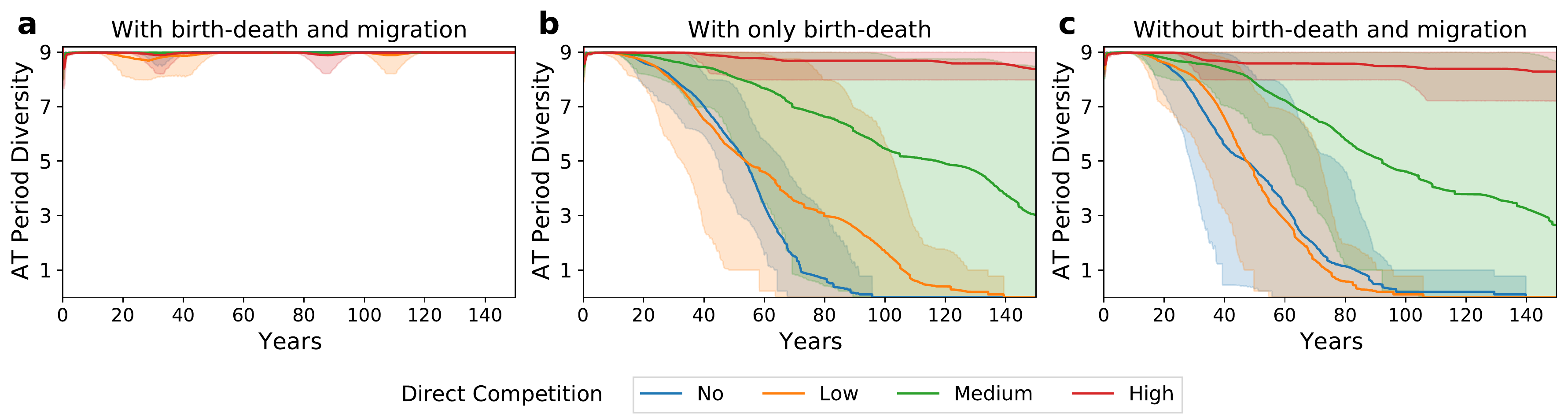}}%
\caption{The inclusion of migration in the model reduces the sensitivity of strain diversity to the assumed level of direct competition. Mean values and $2.5\% - 97.5\%$ quantiles of period diversity of circulating antigenic types (calculated from 10 simulations of the model) are shown when the mean duration of immunity is 30 weeks in demographic settings (a) with birth-death and migration processes, (b) with only birth-death process, (c) without birth-death and migration processes. Levels of direct competition represent scenarios where there are no MTs (no), three MTs (low),  two MTs (medium), and one MT (high). Here, $R_0 \approx 2$ ($c = 10$), $\gamma = 4$ weeks.}
\label{fig:migrationeffect}\end{figure}

Now, we investigate whether our findings are robust to variables in demographic characteristics of the host population. We consider three demographic scenarios: (a) one with birth-death and migration processes, (b) one with only birth-death process, (c) one without birth-death and migration processes (closed population). In Figure \ref{fig:migrationeffect}, we present the endemic period diversity of ATs when the immunity duration is set to 30 weeks for demographic settings (a) - (c). Additional results are presented in the supplementary information \ref{sub.sec.wo.migration} and  \ref{sub.sec.wo.bd.migration}. We find that the sensitivity of endemic diversity to the level of direct competition is similar in the settings with only birth-death process (b), and without birth-death and migration processes (c), indicating that introduction of susceptible people via birth process does not affect the sensitivity of endemic diversity to the level of direct competition. In these settings, higher levels of direct competition lead to higher levels of endemic diversity (Figure \ref{fig:migrationeffect}.b and c). However, in the setting with birth-death and migration processes, there is no observed effect of direct competition on the endemic period diversity, suggesting that the re-introduction of extinct strains through host migration diminishes the sensitivity of endemic diversity to the level of direct competition (Figure \ref{fig:migrationeffect}.a).

\section{Discussion}
For many multi-strain pathogens, we lack a detailed understanding of how competitive interactions within a host affect epidemiological dynamics \citep{alizon2011, bashey2015}. While the indirect competition between strains that arises from host immunity has been well studied using mathematical models \citep{kucharski2016, gog2002,dawes2002,alizon2013}, less attention has been paid to the direct competition between strains that occur within individual hosts. 

In this study, we used an individual-based model to demonstrate how direct competition changes strain diversity.  Compared to the existing models that incorporate direct competition in a phenomenological way, our model enables us to explore a wider range of hypotheses, and link them to observed multi-strain pathogen dynamics. We investigated the effects of direct competition under different assumptions about host immunity, disease transmissibility, and host demographic characteristics. In the following sections, we discuss three contributions of this study. First, we summarise our simulation results which suggest that the relationship between direct and indirect competition mechanisms can affect the expected level of strain diversity. Second, we show how observations of strain diversity of a multi-strain pathogen may be used to infer details of unknown host immune responses. Finally, we show how thorough consideration of the possible mechanisms of direct competition is crucial for models that are used to evaluate future interventions.

\subsection{The strength of indirect competition regulates the impact of direct competition on strain diversity}

Our results illustrate the possibility that outputs of multi-strain models can be highly sensitive to assumptions about direct competition. We found that the degree of sensitivity of endemic strain diversity to these assumptions depends on the strength of direct competition in the model. For example, in scenarios where it is assumed that there is no immunity following infection (no indirect competition), higher assumed levels of direct competition result in lower levels of strain diversity. In scenarios where we assume a lifelong antigen-specific immunity (strong indirect competition), we do not observe any effect of the assumed levels of direct competition on strain diversity due to the strength of indirect competition between strains carrying the same antigenic types that is regulated by the host immune response. However, when we assume an intermediate duration of immunity (15 - 40 weeks), we observe a high sensitivity of diversity to the level of direct competition if $R_0$ is sufficiently low. Otherwise, in higher $R_0$ scenarios, the higher strength of indirect competition (due to higher $R_0$) dominates the competitive dynamics leading to low sensitivity of diversity to the assumed level of direct competition.

\begin{figure}[!htb]
\centering
\makebox[\textwidth][c]{\includegraphics[width=1\textwidth]{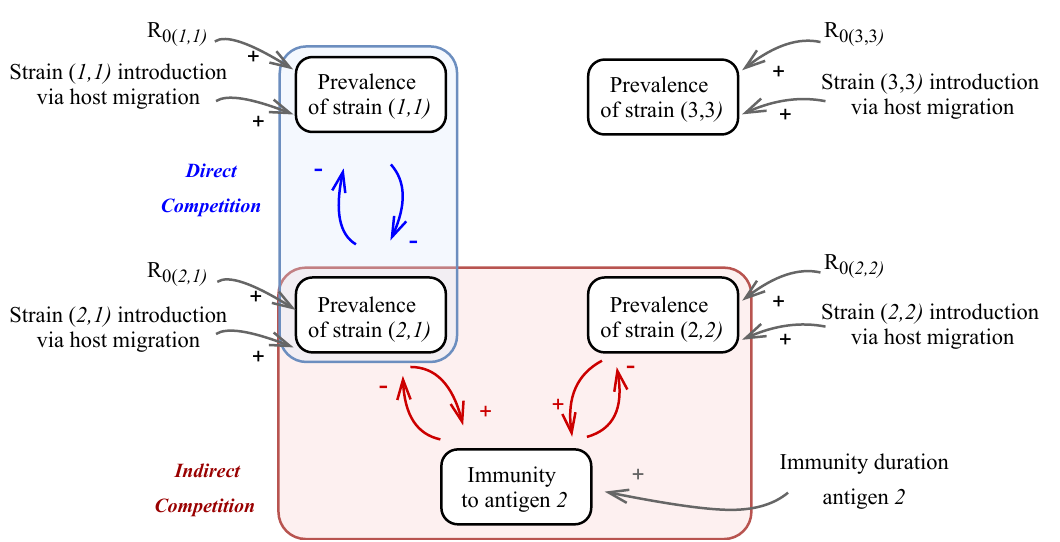}}

\caption{ The strength of direct and indirect competition creates different epidemiological dynamics and changes the strain diversity. An exemplar of underlying mechanisms of observed strain diversity dynamics is shown. Compartments represent population-level prevalence of strains and population-level immunity to antigens. Strains, (\textit{a,m}), are represented by their antigenic types (\textit{a}) and metabolic types (\textit{m}). For simplicity, we only present dynamics with four strains:  (\textit{1, 1}),  (\textit{2, 1}), (\textit{2, 2}), (\textit{3, 3}). In our model, we assume that there is antigen-specific host immune response, and strains with same metabolic types cannot co-infect same host at a given time. 
}
\label{fig:underlying}\end{figure}

We summarise the interaction between direct and indirect competition in our model, and their combined effect on strain prevalence and diversity in Figure \ref{fig:underlying}. The strength of the direct competition (blue positive feedback loop) increases when a higher direct competition or higher $R_0$ are assumed, while the strength of indirect competition (red negative feedback loop) increases when a longer duration of immunity or higher $R_0$ are assumed. The changes in the strengths of these feedback loops lead to different effects on strain diversity. Even though an increase in the strength of either the direct or indirect competition feedback loops tends to decrease the level of strain diversity individually, we observe that a balance between the two feedback loops (\emph{e.g.,} a medium level of immunity duration -40 weeks- and high level of direct competition), can lead to higher levels of strain diversity. We also find that the importation of non-circulating antigenic types via host migration can alter the balance between direct and indirect competition, as it effectively increases the strength of indirect competition.

\subsection{Inferring details of the host immune response to multi-strain pathogens}

Our conceptual framework for understanding the interaction between direct and indirect competition in multi-strain systems (Figure  \ref{fig:underlying}) can help us to interpret the observations of certain multi-strain pathogens. For example, it has been suggested that a Group A \emph{Streptococcus} infection results in an antigen-specific host immune response, however, there is not yet consensus on the duration of immunity after an infection and the strength of within-host competition among strains \citep{pandey2016, hysmith2017, chisholm2020, chisholm2021}. Furthermore, both the prevalence of Group A \emph{Streptococcus} diseases and the diversity of Group A \emph{Streptococcus} strains that are observed in high-income settings are generally lower than those observed in low-income settings \citep{efstratiou2016, smeesters2009}. For example, strain diversity can be up to 5-fold lower in high-income setting, compared to low-income settings  \citep{smeesters2009}. While the $R_0$ of Group A \emph{Streptococcus} in any setting is unknown, it is reasonable to assume that it will be higher in low-income settings where the prevalence of disease is highest. In our model, we observe a substantial increase in strain diversity when $R_0$ is increased from just above 1 in all immunity duration scenarios except those where we assume lifelong immunity (regardless of the assumed level of direct competition) (Figure \ref{fig:atsummaryr0}). Therefore we argue that antigen-specific immunity after Group A \emph{Streptococcus} infection is unlikely to be lifelong. This conclusion supports the results of \cite{chisholm2021} who found that strain-specific immunity acquired by a single infection of duration greater than five years is consistent with epidemiological trends of Group A \emph{Streptococcus}. Further work (\emph{e.g.,} calibrating our model to observations of Group A \emph{Streptococcus} epidemiology) is needed to further narrow down the set of possible durations of immunity for Group A \emph{Streptococcus}.

\subsection{Sensitivity analyses of assumptions about direct competition are crucial if multi-strain models are being used to evaluate non-vaccine intervantions}

\begin{figure}[!htb]
\centering
\includegraphics[width=1\textwidth]{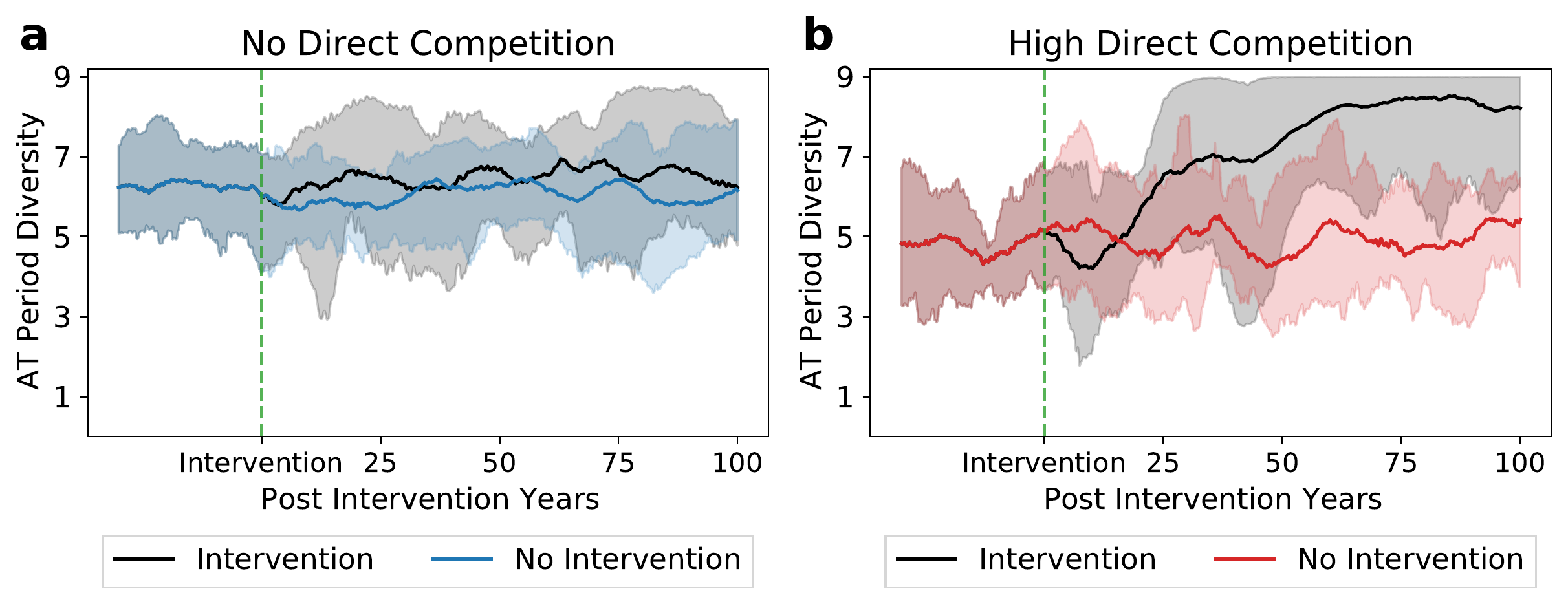}
\caption{An intervention that reduces $R_0$ may either increase or decrease strain diversity, depending on the assumed level of direct competition. Mean values and $2.5\% - 97.5\%$ quantiles of period diversity of circulating antigenic types (from 10 simulations) are shown when there is (a) no direct competition, (b) high direct competition. In no intervention scenarios, $R_0 \approx 4$ ($c = 20$) whereas in the scenarios with intervention, $R_0$ is decreased from $R_0 \approx 4$ ($c = 20$) to $R_0 \approx 2$ ($c = 10$) at the intervention time. Simulations are run for 150 years before intervention. Here, $\mu = 40$ weeks (medium immunity), $\gamma = 4$ weeks, $\alpha = 0.0001$.}
\label{fig:intervention}\end{figure}

Our findings suggest that assumptions about direct competition may influence the estimated impact of an intervention because a lower $R_0$ value does not necessarily correspond to a decrease in strain diversity (Figure \ref{fig:atsummaryr0}). Therefore, interventions designed to reduce disease spread by decreasing disease transmissibility (\emph{e.g.,} \emph{non-vaccine interventions}, such as improving household living conditions or promoting personal hygiene) may not necessarily result in a reduction in strain diversity. To illustrate this point, we considered a non-vaccine intervention designed to decrease $R_0$ by reducing the number of contacts between hosts in the population. In this hypothetical example of a non-vaccine intervention, we assume that the intervention reduces $R_0$ from 4 to 2 at a particular time in the simulation. In the scenarios presented, we assume (a) no and (b) high direct competition (Figure \ref{fig:intervention}). As we expect, the intervention does not change the AT diversity in scenarios that that assume no direct competition, and it increases the AT diversity in scenarios that assume high direct competition. Therefore, reducing $R_0$ with a non-vaccine intervention could either increase or decrease the number of ATs post intervention, depending on the assumed level of direct competition in the model. Understanding which of these scenarios is more likely is crucial in order to effectively evaluate an intervention, particularly if such non-vaccine interventions are followed by, or combined with a multivalent vaccine intervention that targets a subset of circulating strains. Our results illustrate the importance of conducting sufficient sensitivity analyses on the assumed level of of direct competition when using the model to evaluate interventions to reduce the prevalence of multi-strain pathogens.

\subsection{Limitations and future work}
In this study, we used our model to simulate disease transmission within a small population with social connections represented by a fixed heterogeneous network. In settings with smaller populations, differing host network structures may be more appropriate and could lead to different transmission dynamics \citep{buckee2004}. Future work could explore how changing the contact network structure influences our findings. Our study focused on multi-strain pathogens where protective immunity is antigen-specific, and where direct competition is through MT competition. Future modelling work could considered other forms of indirect and direct competition; for example, where it is assumed that (1) hosts develop cross-immunity or have differing durations of immunity after infections with different strains; and/or (2) where there is not only MT competition but also allelopathic interference among strains \citep{bashey2015}. We also assumed that all strains have the same transmissibility; for specific pathogens where some strains are known to be more infectious \citep{watkins2015, alizon2013}, further direct competition analyses could be conducted.

Finally, our study considered a wide range of possible transmission and competition scenarios for a hypothetical multi-strain pathogen and showed that under certain assumptions, model outputs are sensitive to the assumed level of direct competition. This indicate that, for some pathogens, it may be possible to robustly infer the details of the competitive dynamics occurring directly within hosts, by calibrating a model to epidemiological data.

\bibliography{mybibfile}
\newpage
\section*{Supplementary Information}
\beginsupplement

\section{Distribution of Infection and Immunity Durations}
\begin{figure}[!htb]
\centering
\includegraphics[width=1\textwidth]{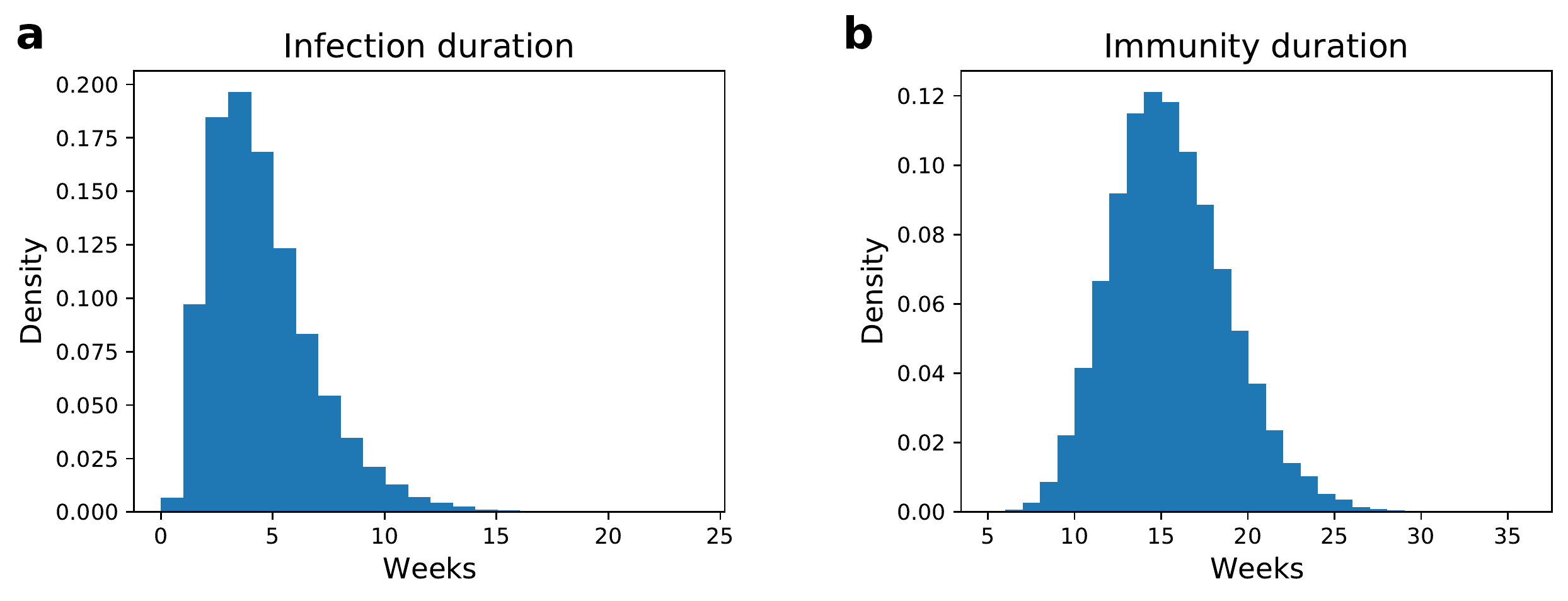}
\caption{Distribution of (a) infection duration with mean duration of 4 weeks, and (b) immunity duration with mean duration of 15 weeks.}
\label{fig:inf_imm_dist}\end{figure}

Infection and immunity durations has gamma distributions with given mean values and shape parameters of  3 and 20, respectively. The high shape parameter of immunity duration allow every recovered person to have immunity for a period of time (Figure \ref{fig:inf_imm_dist} ).

\section{Results without migration} \label{sub.sec.wo.migration}
\begin{figure}[H]
\centering
\includegraphics[width=0.9\textwidth]{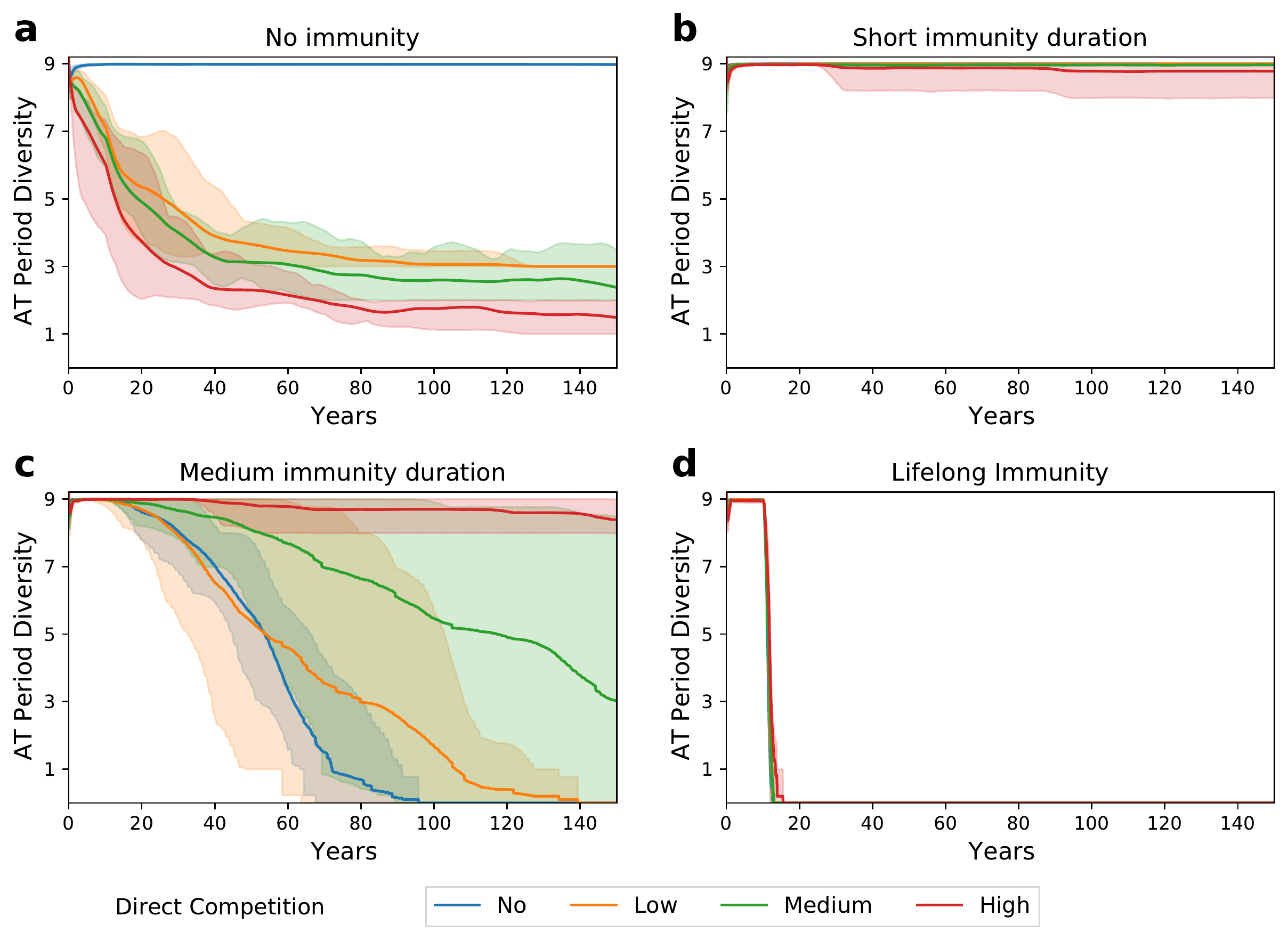}
\caption{Direct competition can increase or decrease strain diversity depending on the duration of immunity. Mean values and $2.5\% - 97.5\%$ quantiles of period diversity of circulating antigenic types under different immunity durations are shown: (a) no immunity duration, (b) short immunity duration (15 weeks), (c) medium immunity duration (30 weeks), (d) lifelong immunity. Levels of direct competition represent scenarios where there are no MTs (no), three MTs (low),  two MTs (medium), and one MT (high). Here, $R_0 \approx 2$ ($c = 10$), $\gamma = 4$ weeks, $\alpha = 0$ (no migration).}
\label{fig:atdiversityBDNoMig}\end{figure}

\begin{figure}[!htb]
\centering
\includegraphics[width=0.9\textwidth]{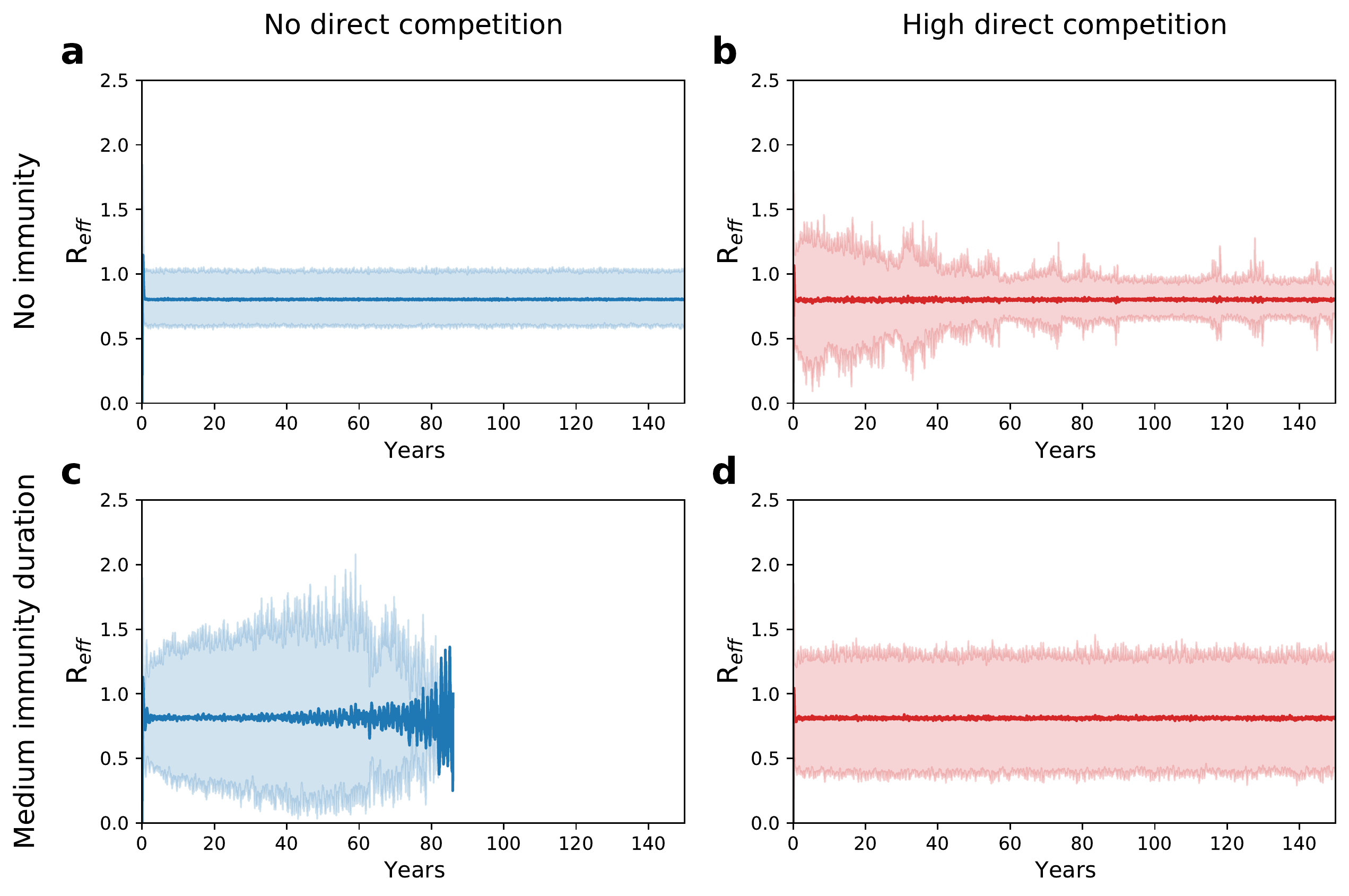}
\caption{The amplitude of oscillations of R\textsubscript{eff} of circulating antigenic types depends on both the level of direct competition and duration of immunity. Mean values and $2.5\% - 97.5\%$ quantiles of R\textsubscript{eff} of circulating antigenic types are shown. First row (a and b) is the results of models with no immunity, second row (c and d) is with medium immunity duration (30 weeks); first column (a and c) is the results of models with no direct competition (blue) and second column (b and d) is the results of models with high direct competition (red). High direct competition represent one MT and no direct competition represents no MTs. Here, $R_0 \approx 2$ ($c = 10$), $\gamma = 4$ weeks, $\alpha = 0$ (no migration).
}
\label{fig:atreffBDNoMig}\end{figure}

The results of population setting without migration (Figure \ref{fig:atdiversityBDNoMig} and \ref{fig:atreffBDNoMig}) are consistent with the one with migration (Figure \ref{fig:atdiversity} and \ref{fig:atreff}). In the setting without migration, we observe a stronger effect of direct competition on strain diversity when there is a medium immunity duration following an infection, \emph{e.g.,} we observe extinction of all the strains in no/low direct competition scenarios when there is a medium immunity duration (Figure \ref{fig:atreffBDNoMig}.c). This suggests that re-introduction of strains via host migration diminishes the effect of direct competition on the observed strain diversity.  When there is lifelong immunity, strains go extinction in all the direct competition level scenarios which means that strain diversity observed in Figure \ref{fig:atdiversity}.d results from host migration.

\clearpage
\section{Results without birth-death and migration processes}  \label{sub.sec.wo.bd.migration}
\begin{figure}[!htb]
\centering
\includegraphics[width=0.9\textwidth]{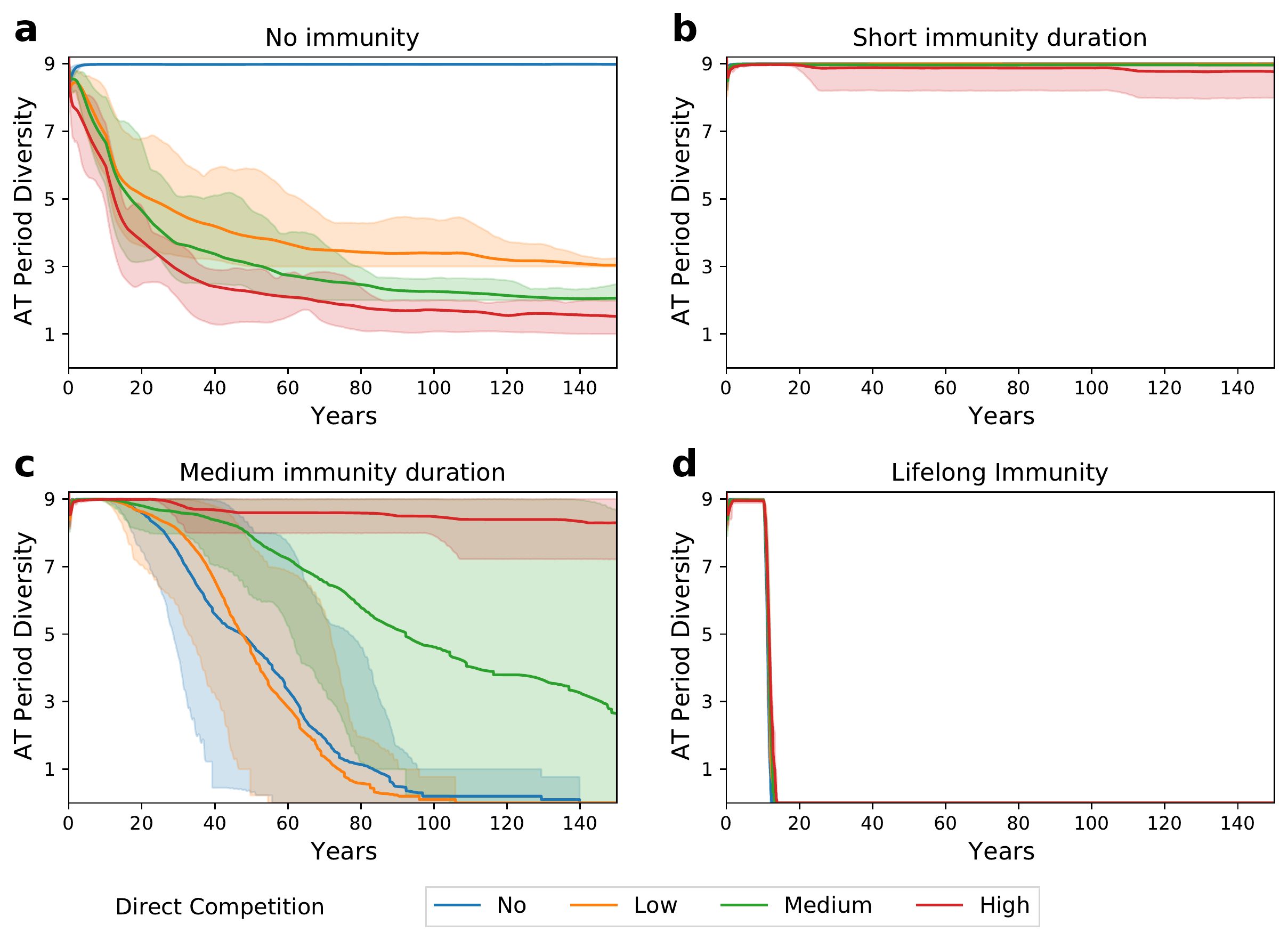}
\caption{Direct competition can increase or decrease strain diversity depending on the duration of immunity. Mean values and $2.5\% - 97.5\%$ quantiles of period diversity of circulating antigenic types under different immunity durations are shown: (a) no immunity duration,  (b) short immunity duration (15 weeks), (c) medium immunity duration (30 weeks), (d) lifelong immunity. Levels of direct competition represent scenarios where there are no MTs (no), three MTs (low),  two MTs (medium), and one MT (high). Here, $R_0 \approx 2$ ($c = 10$), $\gamma = 4$ weeks, $\alpha = 0$ (no migration), there is no birth-death process (closed population).}
\label{fig:atdiversityNoBDNoMig}\end{figure}

\begin{figure}[!htb]
\centering
\includegraphics[width=0.9\textwidth]{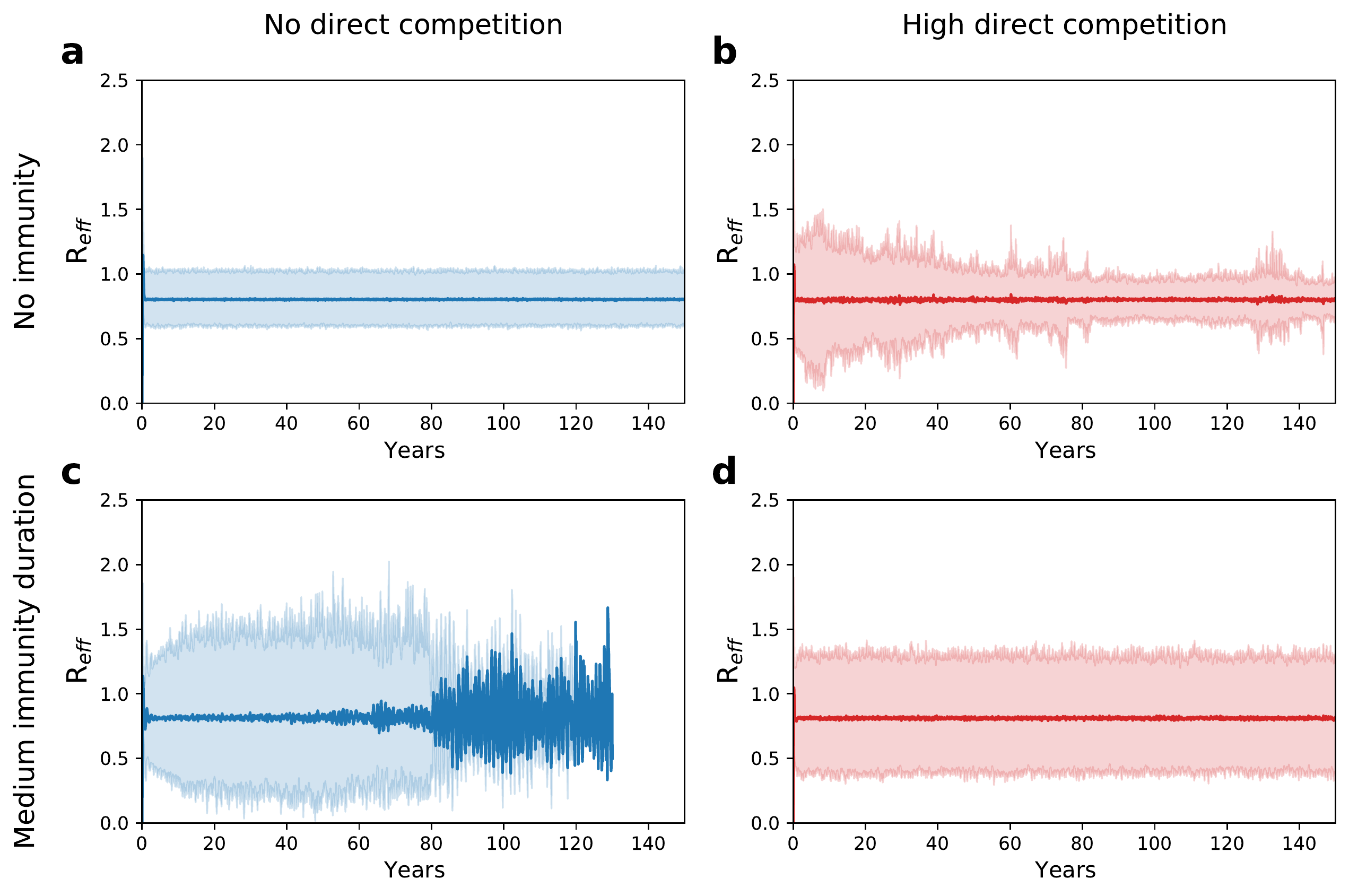}
\caption{The amplitude of oscillations of R\textsubscript{eff} of circulating antigenic types depends on both the level of direct competition and duration of immunity. Mean values and $2.5\% - 97.5\%$ quantiles of R\textsubscript{eff} of circulating antigenic types are shown. First row (a and b) is the results of models with no immunity, second row (c and d) is with medium immunity duration (30 weeks); first column (a and c) is the results of models with no direct competition (blue) and second column (b and d) is the results of models with high direct competition (red). High direct competition represent one MT and no direct competition represents no MTs. Here, $R_0 \approx 2$ ($c = 10$), $\gamma = 4$ weeks, $\alpha = 0$ (no migration), there is no birth-death process (closed population).}
\label{fig:atreffNoBDNoMig}\end{figure}

The results of population setting without birth-death and migration (Figure \ref{fig:atdiversityNoBDNoMig} and \ref{fig:atreffNoBDNoMig}) are consistent with and similar to the one without migration (Figure \ref{fig:atdiversityBDNoMig} and \ref{fig:atreffBDNoMig}). This suggests that introductions of susceptible hosts through birth-death process does not have any significant effect on strain diversity in our parameter space. Compared to the setting with birth-death and migration, in the setting without birth-death and migration, we observe a stronger effect of direct competition on strain diversity when there is a medium immunity duration following an infection. We observe extinction of all the strains in no/low direct competition scenarios when there is a medium immunity duration (Figure \ref{fig:atreffNoBDNoMig}.c). When there is lifelong immunity, strains go extinction in all the direct competition level scenarios.

\end{document}